\documentstyle[12pt]{article}

\begin{document}

\title{A superweak solution of the Strong CP Problem
\footnote{Supported in part by Department of Energy grant
\#DE-FG02-91ER406267}}

\author{{\bf S.M. Barr}\\ 
Bartol Research Institute\\ University of Delaware\\
Newark, DE 19716}

\date{BA-98-36}
\maketitle

\begin{abstract}

A non-axion solution to the Strong CP Problem is proposed that
works even in the context of gravity-mediated supersymmetry breaking.
Both $\epsilon'/\epsilon$ and indirect CP violation in the
$B-\overline{B}$ are predicted to be unobservably small.
$\mu \longrightarrow e \gamma$ is predicted to arise, typically,
with branching ration $3 \times 10^{-12}$. A new source of dark
matter is also predicted in the model.

\end{abstract}

No known solution to the Strong CP Problem seems to be completely
satisfactory. A combination of laboratory, astrophysical, and cosmological
bounds has squeezed invisible axion models into the narrow window
$10^{10}$ GeV $\stackrel{_<}{_\sim} f_a \stackrel{_<}{_\sim} 10^{12}$ 
GeV [1]. Moreover, preserving the low mass of the axion against corrections
from Planck-scale physics is not a simple matter [2]. On 
the other hand, non-axion solutions to the Strong CP Problem have
their own difficulties. The mechanism proposed by Nelson in
[3] does not appear to work in the context of supersymmetry if the
supersymmetry breaking occurs at high scales and is mediated by 
gravitational effects as in
supergravity models, as was pointed out by Dine, Leigh, and Kagan
[4]. A different non-axion mechanism proposed by the present author 
and Zee [5] does not appear to work in supersymmetry no matter how 
supersymmetry is broken. Other solutions to the Strong CP Problem
have been proposed [6,7,8], but none is compelling. Given this state of 
affairs, it is worth looking for other, testable ideas. The idea
that is proposed here has the virtue, besides simplicity,
that it can be implemented in
the context of supersymmetry, whether supersymmetry-breaking is
mediated by gauge or gravitational interactions.
It also leads to interesting and distinctive low-energy 
phenomenology.

As background, it is helpful to understand what goes wrong with
some non-axion solutions to the Strong CP Problem in the context of
supersymmetry. Most of these models are based on the idea that
$\overline{\theta}$ is kept small naturally by a discrete symmetry,
either CP or P, which is spontaneously broken. It can be arranged
in various ways that $\overline{\theta}$ vanishes at tree level and
is induced at one-loop, or sometimes not until two-loop level.
There are two kinds of models of this type. In one type, to which 
the models of [3], [7], and [8] belong, the quark mass matrices have
large CP-violating phases, but have real determinants at tree level.
Such models explain the Kaon-system CP violation using the
standard Kobayashi-Maskawa mechanism. In the other type, to which the 
models of [5] and [6] belong, the quark mass matrices themselves
are real at tree level, and the $\epsilon_K$ parameter arises
from some milliweak or superweak interaction rather than from
the KM phase, which is extremely small.

One of the key problems faced by all such non-axion models
in the context of supersymmetry is that $\overline{\theta}$ 
receives a contribution from the phase of the gluino mass. 
In the models of the type proposed by Nelson [3], this phase
can arise from the one-loop diagram shown in Figure 1(a),
due to a mismatch between the complex quark-mass and squark-mass
matrices.
If supersymmetry breaking occurs at high scales and is mediated
to the observable sector by gravity effects, this one loop
diagram has no natural suppression, and leads to $\overline{\theta}
\sim \alpha_s/4 \pi$, which is many orders of magnitude too large [4].
In the context of gauge-mediated supersymmetry breaking, the
quark-squark mismatch is very small, and the diagram of Fig. 1(a)
can be harmless [9].

In models where the quark mass matrix is real at tree level [5,6],
and where consequently the KM phase is negligible, the Kaon-system
CP violation is often mediated by colored scalar fields. A
characteristic difficulty of these models is that the fermionic 
partners of these fields, being colored fermions, also contribute to 
$\overline{\theta}$, and generally far too much.

The common difficulty of all these models is that the CP-violating 
phases enter into the masses of colored fermions other than the
quarks --- gluinos and/or colored Higgsinos. The question
is whether it is possible to insulate the sector of colored fields more
effectively from the sector in which CP is spontaneously broken, 
and yet have CP violation appear in the Kaon system.

Suppose that in addition to the fields of the Standard Model there
exists a vectorlike pair of quark fields, $D'$ and $D^{'c}$, 
in the $SU(3) \times SU(2)
\times U(1)$ representations $({\bf 3}, {\bf 1}, -\frac{1}{3}) +
(\overline{{\bf 3}}, {\bf 1}, +\frac{1}{3})$. For anomaly freedom
there need to be additional fields, which are most economically
chosen to be lepton doublets, $L'$ and $L^{'c}$, in the representations
$({\bf 1}, {\bf 2}, -\frac{1}{2}) + ({\bf 1}, {\bf 2}, +\frac{1}{2})$.
Thus the new fermions can come from the complete $SU(5)$ multiplets 
${\bf 5} + \overline{{\bf 5}}$. The scalar sector of the Standard Model
is augmented by a pair of complex singlets, denoted $h_1$ and $h_2$. The 
Lagrangian contains the following terms involving the new quark and 
scalar fields:

\begin{equation}
L' = - \sum_{ia} \lambda_{ia} d^c_i D' h_a - M' D^{'c} D' 
- \sum_{ab} m^2_{ab} h^{\dag}_a h_b. 
\end{equation}

\noindent
The masses $M'$ and $m^2_{ab}$ are of order the weak scale.
Note that the terms given in Eq. 1 have a global $U(1)$ symmetry
under which $D'$ has charge $+1$, and $D^{'c}$ and $h_a$ have charge
$-1$. Since the singlet scalars $h_a$ are assumed to have vanishing
vacuum expectation values, this global symmetry, which we shall call
$U(1)_h$, remains unbroken. This symmetry prevents any mixing in the
mass matrix between the new quarks and the known quarks $d$, $s$, and $b$.

It is assumed that CP is a spontaneously broken symmetry, so that
the Lagrangian is CP invariant. Thus, all the quark masses, including
$M'$, are real, as are the Yukawa couplings $\lambda_{ia}$. However,
it is assumed that when CP breaks (at sufficiently high scales that
the resulting domain walls are inflated away) a CP-violating
phase appears in the mass parameter of the singlet scalars, that is
in $m^2_{12} = m_{21}^{2*}$. If we call the fields that break CP spontaneously
$S_A$, then the phase in $m^2_{12}$ could arise through higher-dimension
operators involving factors of $S_A/M_G$. We will
return to this point when we discuss implementing these ideas in 
supersymmetry. Because $\langle h_a \rangle = 0$, the CP-violating 
phase does not appear at tree level in quark masses.

Since the only CP-violating phase at low energy is in the mass of
the singlet scalars $h_a$, the $\epsilon$ parameter of the neutral
Kaon system must arise by means of the box diagram shown in
Fig. 2. In estimating CP-violating effects, we will work in
a basis where the mass matrix of the scalars $h_a$ is real and diagonal,
and where there are consequently phases in the couplings $\lambda_{ia}$.
For simplicity, it will be assumed that the lighter mass eigenstate of 
the singlet scalars (with mass $m$) dominates in loop diagrams. In this 
case, the value of $\epsilon$ is given approximately by

\begin{equation}
\epsilon \cong \left( \frac{B_K m_K f_K^2}{3 \sqrt{2} \Delta m_K}
\right) \frac{1}{192 \pi^2} \frac{{\rm Im}(\lambda_d^{*2} \lambda_s^2)}
{M^2 + m^2} g(m^2/M^2),
\end{equation}

\noindent
where $\lambda_d$ and $\lambda_s$ are the couplings of the lighter 
singlet scalar to $d^c D'$ and $s^c D'$ respectively, and where 
$g(x) = 3x (1 + x) \ln{x}/(1-x)^3 + \frac{3}{2} (1+x)^2/(1-x)^2$.
$g(x)$ is a very slowly varying function, with $g(1) = 1$, and
$g(0) = g(\infty) = 3/2$. Thus, one has that
${\rm Im}(\lambda_d^{*2} \lambda_s^2)/(M^2 + m^2) \approx 
8 \times 10^{-12}$ GeV$^{-2}$. If one assumes that $M \sim m \sim
300$ GeV, and that the CP-violating phase is of order unity, then 
$\left| \lambda_d \lambda_s \right| \sim 10^{-3}$.

In the non-supersymmetric model just described, the leading 
contribution to the strong CP-violating angle $\overline{\theta}$ comes
from the two-loop diagram shown in Fig. 3(a). A crucial point is that
the $h_a$ only couple to the right-handed components of the known
quarks, so that the one-loop diagram of Fig. 3(b) only multiplies the
quark mass matrix by a hermitian factor, and therefore does not contribute 
to $\overline{\theta}$. In Fig. 3(a), the loop factors 
$(16 \pi^2)^{-2}$, and the
Yukawas $\lambda^2 \sim 10^{-3}$, together already give a suppression
of $\overline{\theta}$ by about $10^{-7}$. Given that the quartic
coupling of the term $H^{\prime \dag} H' h^{\dag} h$ can be small,
it is apparent that $\overline{\theta}$ can be sufficiently small
without any fine tuning. 

It is easy to show that the only place where CP violation would
show up in experiments in the foreseeable future is in the parameter
$\epsilon$. $\epsilon'/\epsilon$ must arise also from one-loop
diagrams involving the new particles, and comes out in the range
of $10^{-5}$ to $10^{-6}$. In other words, the predictions are 
essentially those of a superweak model. Similarly, one would expect
no evidence of indirect CP violation to appear in the $B-\overline{B}$
systems. 

However, the model does lead to interesting new physics of other
kinds. One interesting possibility is $\mu \longrightarrow e \gamma$.
By far the simplest way for anomalies to cancel, as we have noted, is
for the new vectorlike quarks to be accompanied by new vectorlike
leptons that fit together with them into the $SU(5)$ multiplets
${\bf 5} + \overline{{\bf 5}}$. In that case, there should also be
couplings of the form $\sum_{ia} \tilde{\lambda}_{ia} L_i L^{'c} h_a +
 \tilde{M} L' L^{'c} + {\rm H.c.}$. 
These allow the one-loop diagrams shown in Fig. 4, which
lead to the following branching ratio for $\mu \longrightarrow e  \gamma$:

\begin{equation}
BR(\mu \longrightarrow e  \gamma) = \frac{\alpha}{768 \pi} 
\frac{\left| \tilde{\lambda}_e \tilde{\lambda}_{\mu} \right|^2}
{G_F^2 (\tilde{M}^2 + m^2)^2} [h(m^2/\tilde{M}^2)]^2,
\end{equation}

\noindent
where $h(x) = (x+1)(x-1)^{-4}(2 x^3 + 3 x^2 -6x +1 -x^2 \ln x)$.
$h(x)$ is a very slowly varying function, with $h(1) = 1$,
$h(0) = 1$, and $h(\infty) = 2$. If one assumes a grand unified
model, and that the ratio
$(\tilde{\lambda}_i/ \tilde{M})/(\lambda_i/M)$ does not run much
in going from the unification scale to the weak scale, then one can
estimate the expression in the last equation by replacing $\tilde{M}$
by $M$ and $\tilde{\lambda}_{e,\mu}$ by $\lambda_{d,s}$. If, then,
$M \sim m \sim 300$ GeV, and $\left| \lambda_d \lambda_s 
\right| \sim 10^{-3}$, as required to get $\epsilon$ to come out right, 
one finds that
$BR(\mu \longrightarrow e  \gamma) \approx 3 \times 10^{-12}$.
Of course, this branching ratio depends on various unknown parameters,
but one can see from this estimate that values only about an order
of magnitude below present bounds are typical.

One would also expect that the new particles in this type of model 
would contribute to the dark matter density of the universe. 
Because the symmetry $U(1)_h$ is unbroken, the lightest particle
carrying its charge is stable. We shall assume that this particle is 
one of the singlet scalars, which we shall call simply $h$. 
The relic abundance of $h$ depends on the annihilation cross
section, which is given by

\begin{equation}
\sigma v \cong (m^2/8\pi)(3 (\sum_i \left| \lambda_i \right|^2)^2
(M^2 + m^2)^{-2}
+ 2 (\sum_i \left| \tilde{\lambda}_i \right|^2)^2)
(\tilde{M}^2 + m^2)^{-2}).
\end{equation}

\noindent
If one assumes that the Yukawas are largest for the third
generation, and that $\tilde{\lambda}_3/\tilde{M} \approx
\lambda_3/M$, then one has that $\sigma v \approx
(5/8 \pi) \left| \lambda_3 \right|^4 m^2/(M^2 + m^2)^2$.
One can then derive a constraint [10] on
the parameters from the requirement that the relic $h$ not 
contribute more than about $0.3$ to $\Omega$. One finds that

\begin{equation}
\lambda_3 \stackrel{_>}{_\sim} M/\sqrt{m (1.7 \times 10^4 {\rm GeV})}.
\end{equation}

\noindent
If $M \sim 300$ GeV, then $\lambda_3 \stackrel{_>}{_\sim} \sqrt{5 
{\rm GeV}/m}$, which is certainly a reasonable value.

The supersymmetrization of this model is straightforward.
The CP-violating mass of the singlet scalars can arise in
the following way. Let the fields whose vacuum expectation values
violate CP spontaneously be gauge singlets called $S_A$.
Suppose that the $h_a$ superfields get supersymmetry-invariant
masses from higher-dimension operators which arise from
integrating out fields at some high scale, such as the unification 
scale $M_G$. These effective operators could be of the form 
$S_A S_B h_a h_b/M_G$, or even higher order in powers of $S_A/M_G$.
If the expectation values of the singlets $S_A$ are less than
$M_G$, these higher-dimension operators can induce masses of
the $h_a$ that are of order the weak scale. The phases of
the expectation values of the $S_A$ will then show up in
the supersymmetric contributions to the off-diagonal mass, 
$m_{12}$ of the $h_a$. The singlets denoted $S_A$ could have 
$U(1)_h$ charge of $+2$, while those denoted $S_B$ could have
vanishing $U(1)_h$ charge. It is simple to construct superpotentials
for the $S_B$ that spontaneously violate CP.

The crucial point is that all the CP violating phases are
appearing in the sector of the gauge-singlet fields $h_a$ and
$S_A$. Therefore, these phases can only show up in the masses
of color fields in a rather indirect way. It is clear, for 
instance, that the one-loop diagram in Fig. 1(a) does not 
contribute to the phase of the gluino mass, since it does not 
involve the singlets scalars. (The fact that the
$h_a$ do not have vacuum expectation values is crucial, as
otherwise the quarks would directly get complex masses from
the terms in Eq. (1).) To get a contribution to the phase of
the gluino mass one must go to the two-loop diagram
shown in Fig. 1(b). This gives a contribution to $\overline{\theta}$
that is of order $\delta \overline{\theta} \sim
(\alpha_s/64 \pi^3) \lambda_3^2 (A m^2/m_{\tilde{g}} M^2)$.
Here $M$ is the mass of the heaviest virtual particle in Fig. 1(b);
$A$ is a typical $A$-parameter of the trilinear SUSY-breaking 
terms; and $m$ is the mass of one of the $h_a$. The loop
factors and Yukawa coupling factors give a number of order
$6 \times 10^{-7}$. The ratios of masses can, for certain
choices of parameters, bring this contribution to $\overline{\theta}$
down to the $10^{-9}$ level. One does not expect, however, that
$\overline{\theta}$ will be much below the present bound.

In a supersymmetric version of this model, there will be
other diagrams contributing to $\epsilon$ and to $\mu \longrightarrow
e \gamma$ besides the ones shown in Fig. 2 and Fig. 4, but
these will not affect the qualitative conclusions of the earlier
discussions.

In conclusion, we have proposed a kind of model in which CP
is spontaneously broken in a gauge-singlet sector that is
somewhat isolated from the sector of colored fields. Some
of these gauge singlets couple ordinary down-type quarks
to new vectorlike down-type quarks. The
CP violation observed in the Kaon system arises from box
diagrams involving the gauge singlet fields and
the new vectorlike quarks. $\epsilon'/\epsilon$ is
unobservably small, so that these models are effectively
superweak in their phenomenology. One also does not
expect to see indirect CP violation in the B systems.
The $\overline{\theta}$ parameter only arises at two loops,
and can be smaller than, but not much smaller than,
$10^{-9}$. The lightest singlet scalar should make a
significant contribution to the dark matter density
of the universe. $\mu \longrightarrow e \gamma$ is expected
at a level not much below $3 \times 10^{-12}$ branching
ratio.

\section*{References}

\begin{enumerate}

\item For a review of axions see R.D. Peccei, in {\it CP Violation},
ed. C. Jarlskog (World Scientific, Singapore, 1989) p 533.
\item M. Kamionkowski and J. March-Russell, {\it Phys. Lett.}
{\bf B 282}, (1992) 137; R. Holman {\it et al.}, {\it Phys.
Rev. Lett.} {\bf 69}, (1992) 1489; S.M. Barr and D. Seckel,
{\it Phys. Rev.} {\bf D46}, (1992) 539. 
\item A. Nelson, {\it Phys. Lett.} {\bf 136B}, (1984) 387;
S.M. Barr, {\it Phys. Rev.} {\bf D30}, (1984) 1805; for a review
see H.Y. Cheng, {\it Phys. Rep.} {\bf 158}, (1988) 1. 
\item M. Dine, R.G. Leigh, and A. Kagan, {\it Phys. Rev.}
{\bf D48}, (1993) 2214.
\item S.M. Barr and A. Zee, {\it Phys. Rev. Lett.} {\bf 55},
(1995) 2253.
\item A. Dannenberg, L. Hall, and L. Randall, {\it Nucl. Phys.}
{\bf B271}, (1986) 574.
\item K.S. Babu and R.N. Mohapatra, {\it Phys. Rev.} {\bf D41}, 
(1990) 1286; S.M. Barr, D. Chang, and G. Senjanovic, {\it Phys. 
Rev. Lett.} {\bf 67}, (1991) 2765; E. Carlson and M.Y. Wang,
Harvard preprint 1992 (unpublished); R.N. Mohapatra, A Rasin,
and G Senjanovic, {\it Phys. Rev. Lett.} {\bf 79}, (1997) 4744.
\item S.M. Barr, {\it Phys. Rev.} {\bf D56}, (1997) 1475;
{\it Phys. Rev.} {\bf D56}, (1997) 5761.
\item S.M. Barr, {\it Phys. Rev.} {\bf D56}, (1997) 1475;
M. Dine, R.G. Leigh, and A. Kagan, {\it Phys. Rev.}
{\bf D48}, (1993) 2214.
\item B.W. Lee and S. Weinberg, {\it Phys. Rev. Lett.}
{\bf 39}, (1977) 165.

\end{enumerate}

\newpage

\noindent
{\bf\large Figure Captions}

\vspace{1cm}

\noindent
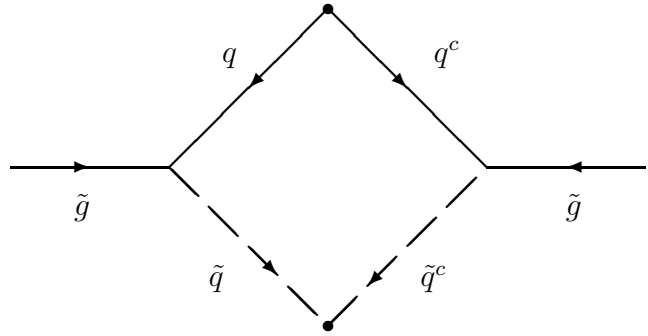
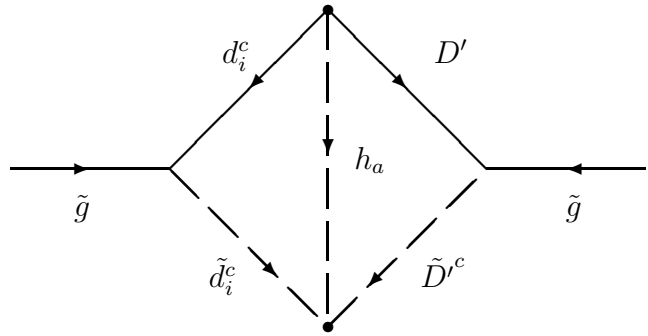
{\bf Fig. 1.} (a) A one-loop diagram that can contribute
to the phase of the gluino mass, and therefore to 
$\overline{\theta}$ in supersymmetric models. (b) In the
model proposed in this paper, the lowest order contribution
to the phase of the gluino mass is this two-loop diagram.

\vspace{0.2cm}

\noindent
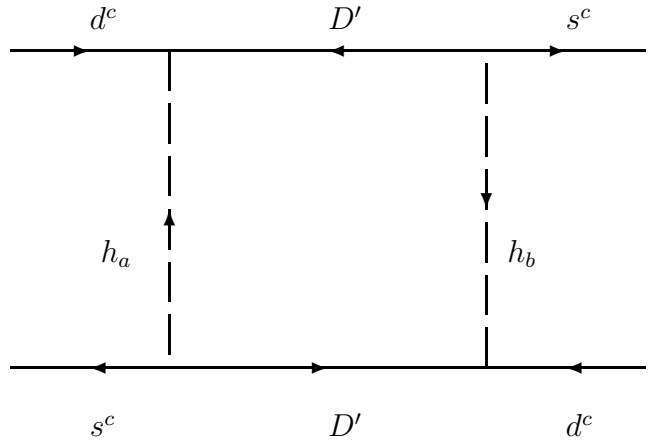
{\bf Fig. 2.} In the proposed model, $\epsilon_K$ arises
from this box diagram. The CP-violating phases appear in
the masses of the gauge-singlet scalars $h_a$.

\vspace{0.2cm}

\noindent
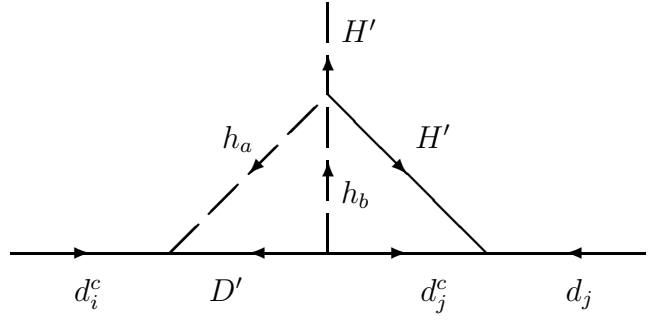
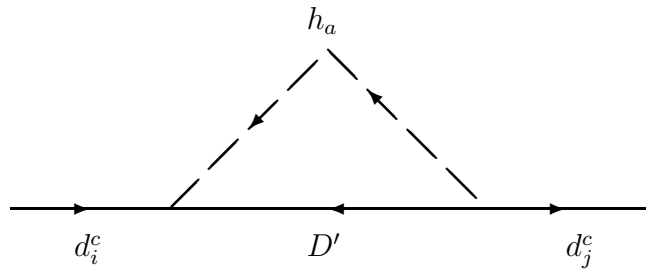
{\bf Fig. 3.} (a) A two-loop contribution to the phase of the
quarm masses. (b) This one-loop diagram does not affect 
$\overline{\theta}$.

\vspace{0.2cm}

\noindent
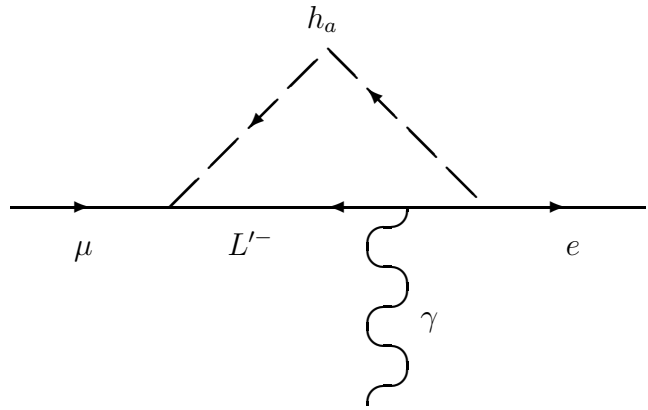
{\bf Fig. 4.} A contribution to $\mu \longrightarrow e \gamma$.

\newpage

\begin{picture}(360,216)
\thicklines
\put(60,108){\vector(1,0){30}}
\put(90,108){\line(1,0){30}}
\put(120,108){\line(1,1){30}}
\put(180,168){\vector(-1,-1){30}}
\put(180,168){\vector(1,-1){30}}
\put(210,138){\line(1,-1){30}}
\put(240,108){\line(1,0){30}}
\put(300,108){\vector(-1,0){30}}
\put(120,108){\line(1,-1){11}}
\put(135,93){\line(1,-1){11}}
\put(150,78){\vector(1,-1){11}}
\put(165,63){\line(1,-1){11}}
\put(180,48){\line(1,1){11}}
\put(206,74){\vector(-1,-1){11}}
\put(210,78){\line(1,1){11}}
\put(225,93){\line(1,1){11}}
\put(180,168){\circle*{4}}
\put(180,48){\circle*{4}}
\put(84,90){$\tilde{g}$}
\put(270,90){$\tilde{g}$}
\put(140,148){$q$}
\put(220,148){$q^c$}
\put(135,63){$\tilde{q}$}
\put(215,63){$\tilde{q}^c$}
\put(160,10){{\bf Fig. 1(a)}}
\end{picture}

\begin{picture}(360,216)
\thicklines
\put(60,108){\vector(1,0){30}}
\put(90,108){\line(1,0){30}}
\put(120,108){\line(1,1){30}}
\put(180,168){\vector(-1,-1){30}}
\put(180,168){\vector(1,-1){30}}
\put(210,138){\line(1,-1){30}}
\put(240,108){\line(1,0){30}}
\put(300,108){\vector(-1,0){30}}
\put(120,108){\line(1,-1){11}}
\put(135,93){\line(1,-1){11}}
\put(150,78){\vector(1,-1){11}}
\put(165,63){\line(1,-1){11}}
\put(180,48){\line(1,1){11}}
\put(206,74){\vector(-1,-1){11}}
\put(210,78){\line(1,1){11}}
\put(225,93){\line(1,1){11}}
\put(180,168){\circle*{4}}
\put(180,48){\circle*{4}}
\put(180,168){\line(0,-1){15}}
\put(180,148){\line(0,-1){15}}
\put(180,128){\vector(0,-1){15}}
\put(180,108){\line(0,-1){15}}
\put(180,88){\line(0,-1){15}}
\put(180,68){\line(0,-1){15}}
\put(190,108){$h_a$}
\put(84,90){$\tilde{g}$}
\put(270,90){$\tilde{g}$}
\put(140,148){$d_i^c$}
\put(220,148){$D'$}
\put(135,63){$\tilde{d}_i^c$}
\put(215,63){$\tilde{D'}^c$}
\put(160,10){{\bf Fig. 1(b)}}
\end{picture}

\newpage 

\begin{picture}(360,216)
\thicklines
\put(60,48){\line(1,0){30}}
\put(120,48){\vector(-1,0){30}}
\put(120,48){\vector(1,0){60}}
\put(180,48){\line(1,0){60}}
\put(240,48){\line(1,0){30}}
\put(300,48){\vector(-1,0){30}}
\put(60,168){\vector(1,0){30}}
\put(90,168){\line(1,0){30}}
\put(120,168){\line(1,0){60}}
\put(240,168){\vector(-1,0){60}}
\put(240,168){\vector(1,0){30}}
\put(270,168){\line(1,0){30}}
\put(120,53){\line(0,1){15}}
\put(120,73){\line(0,1){15}}
\put(120,93){\vector(0,1){15}}
\put(120,113){\line(0,1){15}}
\put(120,133){\line(0,1){15}}
\put(120,153){\line(0,1){15}}
\put(240,163){\line(0,-1){15}}
\put(240,143){\line(0,-1){15}}
\put(240,123){\vector(0,-1){15}}
\put(240,103){\line(0,-1){15}}
\put(240,83){\line(0,-1){15}}
\put(240,63){\line(0,-1){15}}
\put(90,176){$d^c$}
\put(180,176){$D'$}
\put(270,176){$s^c$}
\put(90,22){$s^c$}
\put(180,22){$D'$}
\put(270,22){$d^c$}
\put(94,88){$h_a$}
\put(248,88){$h_b$}
\put(160,0){{\bf Fig. 2}}
\end{picture}

\newpage

\begin{picture}(360,216)
\thicklines
\put(60,108){\vector(1,0){30}}
\put(90,108){\line(1,0){30}}
\put(120,108){\line(1,1){11}}
\put(135,123){\line(1,1){11}}
\put(161,149){\vector(-1,-1){11}}
\put(165,153){\line(1,1){11}}
\put(180,168){\vector(1,-1){30}}
\put(210,138){\line(1,-1){30}}
\put(240,108){\line(1,0){30}}
\put(300,108){\vector(-1,0){30}}
\put(120,108){\line(1,0){30}}
\put(180,108){\vector(-1,0){30}}
\put(180,108){\vector(1,0){30}}
\put(210,108){\line(1,0){30}}
\put(180,168){\vector(0,1){15}}
\put(180,188){\line(0,1){15}}
\put(180,148){\line(0,1){15}}
\put(180,128){\vector(0,1){15}}
\put(180,108){\line(0,1){15}}
\put(185,127){$h_b$}
\put(84,90){$d_i^c$}
\put(270,90){$d_j$}
\put(140,148){$h_a$}
\put(213,148){$H'$}
\put(135,90){$D'$}
\put(215,90){$d_j^c$}
\put(185,188){$H'$}
\put(160,46){{\bf Fig. 3(a)}}
\end{picture}

\begin{picture}(360,216)
\thicklines
\put(60,108){\vector(1,0){30}}
\put(90,108){\line(1,0){30}}
\put(120,108){\line(1,1){11}}
\put(135,123){\line(1,1){11}}
\put(161,149){\vector(-1,-1){11}}
\put(165,153){\line(1,1){11}}
\put(180,168){\line(1,-1){11}}
\put(206,142){\vector(-1,1){11}}
\put(210,138){\line(1,-1){11}}
\put(225,123){\line(1,-1){11}}
\put(240,108){\vector(1,0){30}}
\put(270,108){\line(1,0){30}}
\put(120,108){\line(1,0){60}}
\put(240,108){\vector(-1,0){60}}
\put(84,90){$d_i^c$}
\put(270,90){$d_j^c$}
\put(172,176){$h_a$}
\put(172,90){$D'$}
\put(160,46){{\bf Fig. 3(b)}}
\end{picture}

\newpage

\begin{picture}(360,216)
\thicklines
\put(60,108){\vector(1,0){30}}
\put(90,108){\line(1,0){30}}
\put(120,108){\line(1,1){11}}
\put(135,123){\line(1,1){11}}
\put(161,149){\vector(-1,-1){11}}
\put(165,153){\line(1,1){11}}
\put(180,168){\line(1,-1){11}}
\put(206,142){\vector(-1,1){11}}
\put(210,138){\line(1,-1){11}}
\put(225,123){\line(1,-1){11}}
\put(240,108){\vector(1,0){30}}
\put(270,108){\line(1,0){30}}
\put(120,108){\line(1,0){60}}
\put(240,108){\vector(-1,0){60}}
\put(84,90){$\mu$}
\put(270,90){$e$}
\put(172,176){$h_a$}
\put(142,90){$L^{\prime -}$}
\put(160,10){{\bf Fig. 4}}
\put(202.5,108){\oval(15,15)[br]}
\put(202.5,93){\oval(15,15)[l]}
\put(202.5,78){\oval(15,15)[r]}
\put(202.5,63){\oval(15,15)[l]}
\put(202.5,48){\oval(15,15)[r]}
\put(202.5,33){\oval(15,15)[tl]}
\put(215,63){$\gamma$}
\end{picture}

\end{document}